\documentclass[final,times, 5p]{elsarticle}
\usepackage{amsmath}
\usepackage{graphicx}
\usepackage{float}
\usepackage{xcolor}
\usepackage{graphicx}
\usepackage{dcolumn}
\usepackage{bm}
\usepackage{lineno}
\usepackage{multicol}

\usepackage[T1]{fontenc}

\def \etal{{\it et. al}}

\begin{document}

\title{Real-time sextupole tuning for a long in-plane polarization at storage rings}

\author{Selcuk Hac{\i}{\" o}mero{\u g}lu\corref{cor1}}
\cortext[cor1]{selcuk.haciomeroglu@gmail.com}
\address{Center for Axion and Precision Physics Research, IBS, Daejeon 34051, Republic of Korea}

\begin{abstract}
	A long in-plane beam polarization can be a desired feature for spin measurement experiments in storage rings. The spin precession of the particles within a beam can be controlled by means of the frozen spin method and beam bunching via RF cavities, eventually yielding a polarization lifetime of 10--100 seconds. Previous studies have shown that it can be further improved by sextupoles, which correct the second order effects related to the chromaticity of the beam. However, sextupoles can require readjustment after slight changes in ring parameters. This work presents a real-time sextupole tuning method that relies on a feedback algorithm. It adjusts the sextupole strength during storage, targeting a zero average radial spin component. Satisfying this condition results in a longer polarization lifetime. Simulation studies show that roughly determined feedback coefficients in this method work effectively for a wide range of ring parameters, with practical field imperfections and measurement errors taken into account. Alternatively, this technique can be used to optimize sextupole strengths in a test run without intervening the measurement.
\end{abstract}


\maketitle

 
\section{Introduction}
Storage ring spin physics experiments may require long polarization lifetime for better sensitivities. Previous studies showed that a slow spin precession, which is required for a long polarization lifetime can be achieved by means of ``frozen spin method''  \cite{ref:edm_in_sring, ref:magic_mom_arxiv}, beam bunching via RF cavities \cite{ref:nima_rk4}, and sextupoles \cite{ref:elec_sext,ref:cosy_sext_1,ref:cosy_sext_2}. The optimization procedure in the last method is a time consuming process, especially when the experimental conditions vary at different injections.

This work investigates the feasibility of a real-time sextupole tuning method through beam and spin dynamics simulations. The method relies on periodic measurements of beam polarization and generating a feedback sextupole field to minimize it. Section \ref{sect:spindyn} introduces the basic concepts and methods regarding in-plane spin polarization. Section \ref{sect:fixedsext} shows an application of sextupoles for a long polarization lifetime, and addresses the adjustment requirements. Real-time sextupole tuning is introduced and its performance for various beam configurations is discussed in Section \ref{sect:varsext}. Finally, several cases regarding unwanted fields and imperfect measurements are discussed in Section \ref{sect:imperfections}.

\section{Spin dynamics in a storage ring}\label {sect:spindyn}
 The spin precession in a storage ring is determined by the T-BMT equation \cite{ref:bmt}
\begin{equation}
\small
\begin{split}
\frac{d \vec s}{dt} = -\frac{e}{m}\vec s \times & \left[\frac{G \gamma+ 1}{\gamma} \vec B - \frac{G \gamma}{\gamma+1}\vec \beta (\vec \beta \cdot \vec B) \right. -\left(G+\frac{1}{\gamma+1} \right)\frac{\vec \beta \times \vec E}{c} \\
&+  \left.\frac{\eta}{2c}\left(\vec E -\frac{\gamma}{\gamma+1} \vec \beta (\vec \beta \cdot \vec E) + c \vec \beta \times \vec B \right) \right] ,
\end{split}
\label{eq:bmt}
\end{equation}
where $c$, $e$ and $m$ are the speed of light, the electric charge and mass of the particle, $G=g/2-1$ is the magnetic anomaly ($\approx 1.8$ for proton),  $\vec \beta$ and $\gamma$ are the relativistic velocity and the Lorentz factor, $\vec B$ and $\vec E$ are the magnetic and electric fields respectively. In the ideal case, the average magnetic field on the beam is zero, and the electric field is purely radial. The EDM coefficient $\eta$ is approximately $2 \times 10^{-15}$ for proton EDM of $d=10^{-29} ~e\cdot$cm.

It would be instructive to explain some concepts by referring to an experiment. Storage ring electric dipole moment (EDM) experiments rely on measurements of out-of-plane spin precession rate $d \vec s_y/~dt$ of a longitudinally polarized charged beam \cite{ref:edm_in_sring, ref:edm_paper, ref:deuteron_proposal}. Assuming a total cancellation of the magnetic field, it precesses as 
\begin{equation}
\frac{d \vec s_y}{dt}=-\frac{e\eta}{2mc}\vec s \times \left(\vec E -\frac{\gamma}{\gamma+1} \vec \beta (\vec \beta \cdot \vec E) \right) .
\label{eq:dsy_dt}
\end{equation}
On the other hand, the in-plane (i.e. horizontal, towards the ring center) spin precession rate with respect to the momentum is obtained by subtracting the cyclotron motion from Eq. \ref{eq:bmt}:
\begin{equation}
\frac{d \vec s_x}{dt}=\frac{e}{m}\vec s \times \left( G-\frac{1}{\gamma^2-1} \right)\frac{\vec \beta \times \vec E}{c}.
\label{eq:dsx_dt}
\end{equation} 
Comparison of $G$ and $\eta$ shows that the out-of-plane spin precession tends to be orders of magnitude slower than the in-plane precession. Unless the latter is ``frozen'', and prevented from oscillating, the out-of-plane spin component cannot grow with time because the main signal is proportional to the product of longitudinal spin and radial electric field components as shown in Eq. \ref{eq:dsy_dt}. It is possible to work with a fast in-plane spin precession (at a frequency $f_G$) and measure the oscillation amplitude of the out-of-plane spin component at that specific frequency. This method was applied in the muon EDM measurement \cite{ref:muonEDM}. However, the sensitivity dropped by orders of magnitude in this method.

Limiting the in-plane spin precession helps reducing several systematic errors as well. One of them is a false EDM signal, originating from the coupling between $G$ and the longitudinal magnetic field, as seen in the first and the second terms of Eq. \ref{eq:bmt}. This effect is enhanced proportionally with $s_x$ \cite{ref:b_field_paper}. Another false EDM signal originates from the coupling between $G$, and the product of the vertical component of $\vec \beta$ and the radial component of the electric field ($3^\text{rd}$ term in Eq. \ref{eq:bmt}). Again, it is enhanced by $s_x$ \cite{ref:symm_hybrid_paper}.  

The in-plane spin precession rate can be ideally stopped by the correct choice of $\gamma=\sqrt{1/G+1}$, corresponding to the magic momentum $p_0$ ($\approx 0.7$ GeV/$c$ for proton). This is the ``frozen spin method''. However, momentum spread of the beam limits the performance of the frozen spin method because the off-momentum ($\Delta p/p_0 \ne 0$) particles in the beam have non-zero in-plane spin precession. This causes a decoherence among the particles, and the average vertical spin component stops accumulating. Following this argument, the spin coherence time (SCT) is defined as the time needed for keeping the off-momentum populations' spin spread within 1 radian.

An off-momenta particle orbits the storage ring with an average horizontal offset $\Delta x$. An RF cavity can make it oscillate around $\Delta x = 0$, which eventually prolongs the SCT \cite{ref:nima_rk4}. To first order, this is equivalent to averaging its momentum to $p_0$. However, second order effects, originating from momentum spread and free betatron oscillations still cause a depolarization. 

Regarding these second order effects, it was previously shown that for electron/positron beams the beam polarization could be prolonged by minimizing the chromaticity by means of sextupoles \cite{ref:elec_sext}. Y. Orlov has made first analytical estimations of the sextupole strengths for the deuteron beam to achieve a SCT of order $10^3$ seconds \cite{ref:yuri_sext}. 

At the COSY magnetic storage ring, the SCT of the deuteron beam could be prolonged by means of beam bunching with an RF cavity and optimizing three sextupole families, which were located where three functions, $\beta_x$, $\beta_y$ and the dispersion obtained the maximum values. A SCT at the order of $10^3$ seconds was achieved for various sextupole configurations. Based on an extrapolation, they reported that longer SCT was also achievable \cite{ref:cosy_sext_1}. The sextupoles in those tests were first optimized by scanning, then the chromaticity and $s_x$ were monitored. They have observed a linear relation between the sextupole strength and the absolute value of the polarization loss.  The slope of the line was reproducible at different runs. However, the zero crossing, an indication of the largest SCT depended on the experimental setup. That means, the sextupole configuration requires a separate adjustment for different experimental conditions. A detailed analysis of the sextupole optimization and its connection to chromaticity is given in Ref. \cite{ref:cosy_sext_2}.

\section{Achieving longer spin coherence with sextupoles}\label {sect:fixedsext}

\begin{figure}
	\centering
	\includegraphics[width=\linewidth]{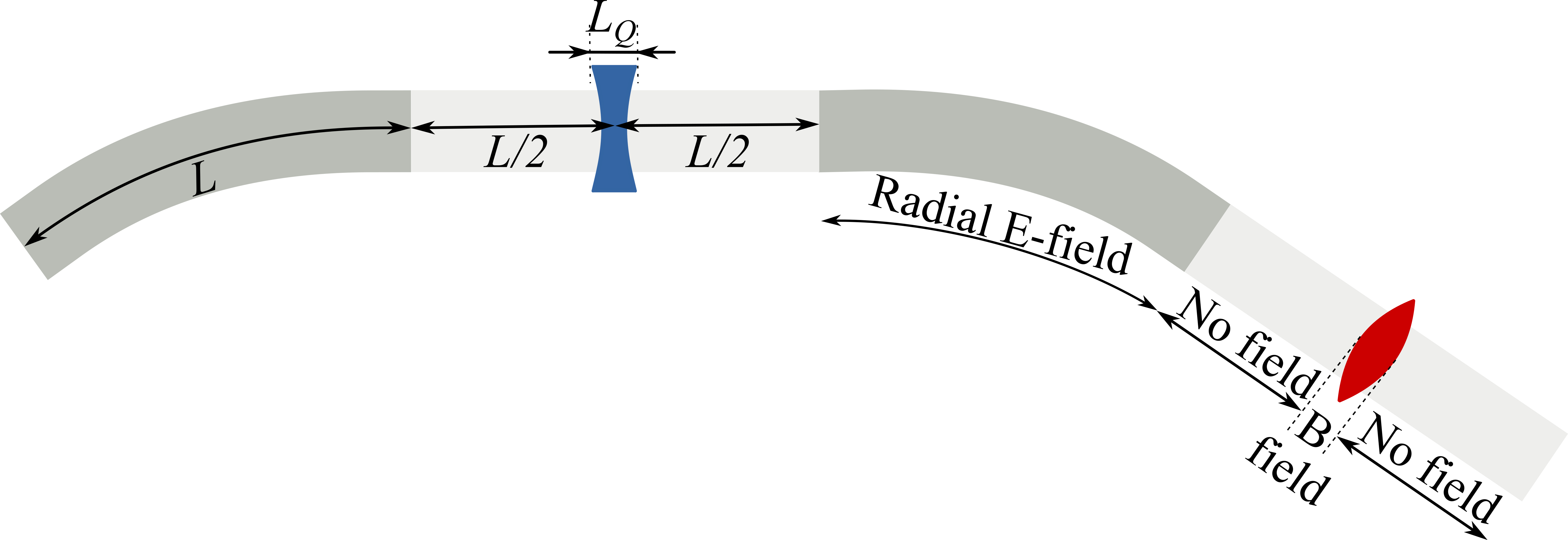}
	\caption{Unscaled depiction of a cell in the symmetric hybrid storage ring. The red and blue lenses correspond to focusing and defocusing quadrupoles. The same alternating polarity pattern is also applied to the  sextupoles, overlapping with the quadrupoles. They are located at the centers of the straight sections, which would otherwise be free of electric and magnetic fields. The bending sections provide a purely radial electric field. The length of a deflector is equal to the length of a straight section. 48 identical cells of $L=7.25$ m and $L_Q=40$ cm comprise a circumference of 696 m. Also an RF cavity is located at the center of one of the straight sections.}
	\label{fig:symmetric_ring}
\end{figure}

The simulations in this work were conducted for the hybrid storage ring design for the proton EDM experiment \cite{ref:symm_hybrid_paper}, with a simple lattice as shown in Figure \ref{fig:symmetric_ring}. However, the method is applicable to other storage rings as well. The term ``hybrid'' refers to the combination of electric deflectors and magnetic quadrupoles. Every quadrupole has the same focusing strength with an alternating polarity pattern: the beam sees focusing and defocusing quadrupoles consecutively at every straight section. 

The sextupole elements in these studies overlapped with the 40 cm long quadrupoles, which were located at the center of every straight section. The sextupole strength $k_s$ was defined as 
\begin{equation*}
B_x = -2k_s x y~;~~~
B_y = -k_s (x^2-y^2),
\end{equation*}
where $B_x$ and $B_y$ are the magnetic flux density components, $x$ and $y$ are the horizontal and vertical positions of the beam respectively. Similar to the quadupoles, each sextupole had a fixed absolute strength with an alternating polarity pattern. A number of simulations were repeated with different $k_s$ values, and then $k_s$ was optimized for the smallest change of rate in $s_x$. The RF cavity was enabled at every simulation in this work. 

\begin{figure}
	\centering
	\includegraphics[width=1\linewidth]{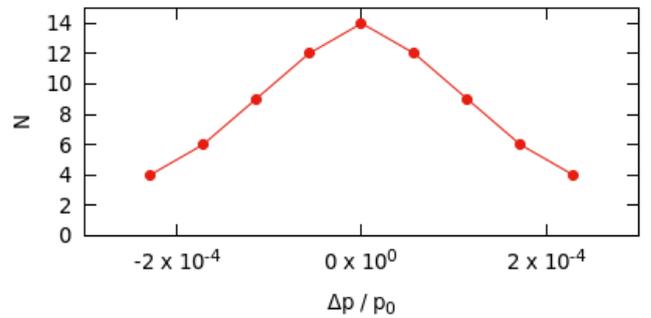}
	\caption{Momentum distribution of the 74 particles used in the simulations. The circles serve for guiding the eye. The particles are distributed along the curve. The distribution is approximately a Gaussian of $\sigma=\Delta p/p_0=10^{-4}$ with the tails truncated beyond 2$\sigma$.} 
	\label{fig:distribution}
\end{figure}

\begin{figure}
	\centering
	\includegraphics[width=1\linewidth]{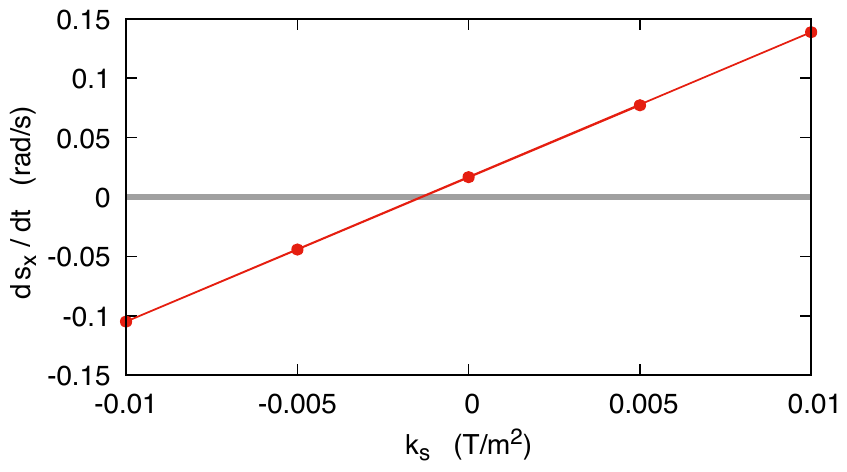}
	\caption{Spin precession rate  $ds_x/dt$ vs. sextupole strength, based on the simulations using a beam of 74 particles. The zero crossing is at $k_s\approx -0.00134$ T/m$^2$.}
	\label{fig:sx_vs_ks}
\end{figure}

The beam in the simulations had 74 particles with an approximately Gaussian distribution (Figure \ref{fig:distribution}), where the particles with $|\Delta p/p_0|>2 \times 10^{-4}$ were excluded. The particles were simulated simultaneously. The spin precession of the whole population was estimated by
\begin{equation}
s_x(t) = \frac{1}{N}\sum_{i=1}^{N}{s_x^i(t)},
\label{eq:avg_sx}
\end{equation}
where $t$ is the simulation time, $s_x^i(t)$ is the horizontal spin component of the $i^\text{th}$ particle and $N=74$ is the number of the simulated particles. Figure \ref{fig:sx_vs_ks} shows the average in-plane spin precession rate $ds_x/dt$ of the beam as a function of the sextupole strength $k_s$. A linear fit yields $k_s \approx -0.00134$ T/m$^2$ for the zero crossing. 

\begin{figure}
	\centering
	\includegraphics[width=1\linewidth]{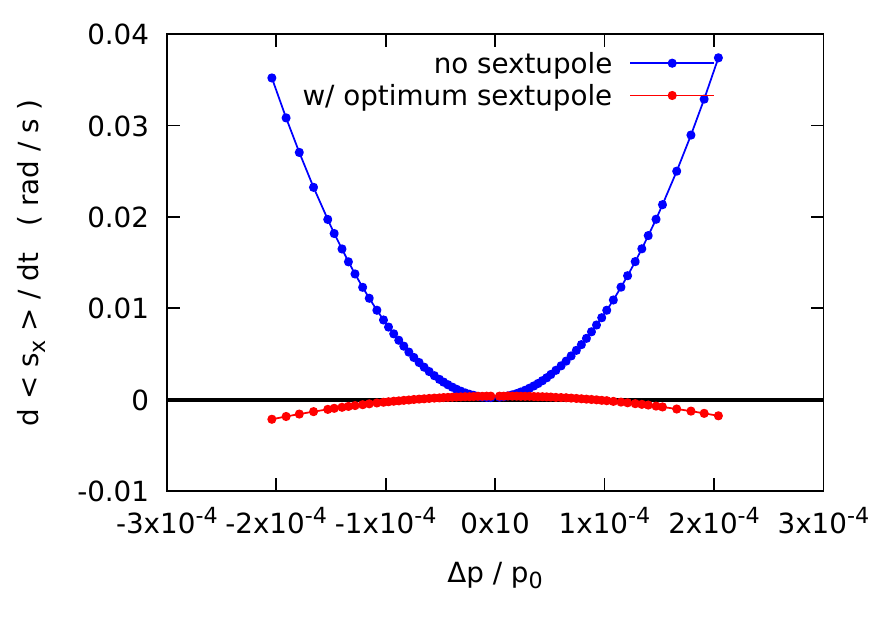}
	\caption{Spin precession rate as a function of momentum spread $\Delta p/p_0$. The sextupoles with $k_s=-0.00134$ T/m$^2$  improved the spin coherence significantly. }
	\label{fig:sct_vs_dp_p_no_pid}
\end{figure}

Figure \ref{fig:sct_vs_dp_p_no_pid} shows simulation results of $ds_x/dt$ as a function of momentum spread. The overall spin coherence was improved significantly with ($k_s=-0.00134$ T/m$^2$) sextupoles. A comparison of Figure \ref{fig:sx_vs_ks} and Figure \ref{fig:sct_vs_dp_p_no_pid} shows that zero crossing corresponds to a full cancellation between positive and negative spin precession rates, rather than an infinite SCT. That is, it can be observed even with highly depolarized beams. 

\begin{figure}
	\centering
	\includegraphics[width=0.9\linewidth]{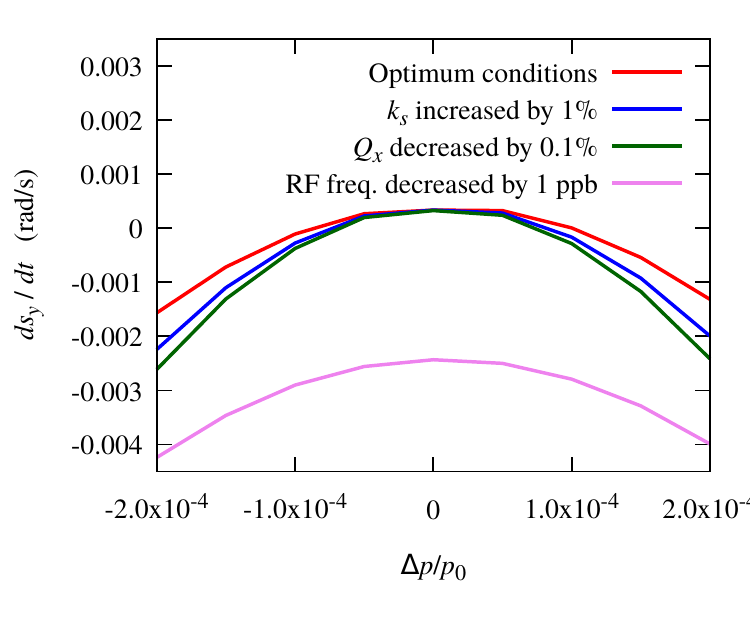}
	\caption{Variations in quadrupole and sextupole strengths can result in spin decoherence. Comparisons with the optimum condition are given in Table \ref{tbl:spr_vs_shifts}.  In case of an RF frequency variation, first order effects prevail in spin precession. The plot shows that the RF frequency should be stable at 1 ppb level or better for the EDM experiments.}
	\label{fig:shifts_with_fixed_sextupole}
\end{figure}

\begin{table*}
	\caption{Spin precession rate changes as the fields deviate from the optimum conditions for fixed value sextupoles. The left column describes the change in the applied field with respect to the optimum conditions. The other columns show how the spin precession rate of various momenta populations changes.}
	\centering
	\begin{tabular}{c c c c}
		\hline
		Change in the field&\multicolumn{3}{l} {Spin precession rate with respect to optimum conditions} \\
		& $\Delta p/p_0=0$ &  $\Delta p/p_0=10^{-4}$ & $\Delta p/p_0=2 \times 10^{-4}$ \\
		\hline
		1\% increase in $k_s$ & negligible &$-0.2$ mrad/s & $-0.7$ mrad/s \\
		0.1\% decrease in $Q_x$ &negligible & $-0.3$ mrad/s & $-1$ mrad/s \\
		\hline

	\end{tabular} 
	\label{tbl:spr_vs_shifts}
\end{table*}

While an optimum sextupole configuration improves spin coherence, it should be tuned extremely well. Moreover, changes in other fields can have significant impact on the spin precession. A number of studies were conducted to determine the effect of the changes in RF frequency, and quadrupole and sextupole strengths. The results are shown in Figure \ref{fig:shifts_with_fixed_sextupole} and Table \ref{tbl:spr_vs_shifts}. The simulations showed that a 1\% drift in $k_s$ increased the spin precession rate of nonzero $\Delta p/p_0$ momenta populations by up to 50\%. A larger effect was observed when the quadrupole strength was decreased by 0.1\%. It will be shown below that the real-time tuning method performs significantly better against changes in the quadrupole strength. Being a first order effect, a 1 ppb RF frequency shift had a large effect on the spin precession too. All the particles precessed with a much higher rate than the optimum case. It is clear that regardless of the sextupoles, in order to achieve a frozen spin, the RF frequency should be tuned to 1 ppb accuracy or better. 

Besides the stability issues, varying ring parameters at different runs may be a preferred feature as in the hybrid ring design for the proton EDM experiment \cite{ref:hybrid}. In that case, the sextupoles need to be adjusted for each experimental configuration. Along with simplifying this process significantly, a real-time sextupole tuning provides a one-size-fits-all solution to the sextupole adjustment problem in general.

\section{Real-time sextupole tuning}\label {sect:varsext}
The previous section showed that spin coherence could be improved significantly through a precise tuning of $k_s$, and that the performance is sensitive to field drifts. It will be shown in this section that $k_s$ can be tuned automatically during storage time using a feedback system, namely a PID controller. 

PID stands for (P)roportional-(I)ntegral-(D)erivative, representing the three operations used for calculating the feedback. The response $R$ of the PID controller is given as
\begin{equation}
R = K_P\epsilon + K_I\epsilon^\text{tot} + K_D\epsilon^\text{diff},
\label{eq:pid_formula}
\end{equation}
where the error function $\epsilon$ is the difference between the desired and obtained outputs,  the superscripts "tot" and "diff" refer to integral over time, and change at the last step, respectively. The feedback coefficients have different functionalities:
\begin{itemize}
	\item  The proportional coefficient $K_P$ weighs the error $\epsilon$. If the controller has only this term, the output oscillates nearby the target value with an offset.
	\item  The integral coefficient $K_I$ weighs the integral of the past error to reduce the offset from the first term. A large $K_I$ value results in unstable oscillations.
	\item The derivative coefficient $K_D$ weighs the difference between the last two measurements. It is used for damping the oscillations and converging quickly.  
\end{itemize}

As shown in Figure \ref{fig:sct_vs_dp_p_no_pid}, the spin precession rate has a quadratic shape with respect to the momentum spread. Changing the sextupole field dominantly precesses the spin of each population in the same direction, yet at different rates. Therefore, the total in-plane spin component of the beam ($s_x$) and its time derivative ($ds_x/dt$) approach zero simultaneously. This behavior lets the time average $\langle s_x \rangle$ be used as an error function, which is preferable because of being relatively easy to measure.

Finally, this method does not require an initial prediction about the optimum sextupole strength. It only requires a coarse adjustment of the feedback coefficients, which do not need to be changed unless the ring parameters are dramatically changed. Then, the sextupole tuning process is taken over by the simple PID controller to converge $k_s$ to the optimum value during the storage.

\subsection{Adjusting the PID parameters}
The simulations for the parameter adjustment were conducted with a particle with $\Delta p/p_0=10^{-4}$. The  time average $\langle s_x \rangle$ is defined as the error function and measured periodically at every 0.1 ms. This period is large enough to prevent the PID controller from reacting the betatron oscillations. After each calculation, $k_s$ was updated by Eq. \ref{eq:pid_formula}.

The coefficients $K_P,~K_I$, and $K_D$ can be adjusted in a number of ways. In this work, they were determined by manual tests, one by one. Initially, $K_I$ and $K_D$ were set to zero, and $K_P$ was increased until $\langle s_x \rangle$ started to make visible oscillations in the desired time scale. At that state, the spin precession oscillated with an offset, while $k_s$ oscillated around the optimal value. 

Then, $K_P$ was kept at that value and new simulations were made by increasing $K_I$ values until the offset in $\langle s_x \rangle$ was corrected. However, some time after damping, $\langle s_x \rangle$ started oscillating in an unstable way. This seemed to be mostly related to the fact that $K_I$ was trying to fix the large synchrotron oscillations. Probably increasing the measurement from 0.1 ms to a second would fix this problem, but it was not feasible for a ms scale simulation. Therefore, $K_I$ was set to zero in the rest of the simulations. It is worth noting that the offset is not too critical as long as the spin precession rate is concerned.

Finally, several simulations were made by increasing $K_D$ until the oscillations were observed to decay sufficiently fast. The PID coefficients did not need to be fine-tuned for different experimental conditions, while it might be helpful for a faster convergence. They had fixed values during each simulation.

\begin{figure}
	\centering
	\includegraphics[width=1\linewidth]{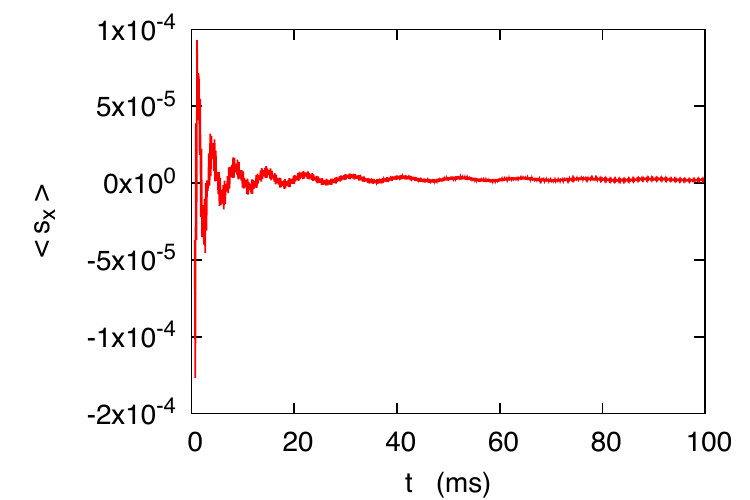}
	\caption{Time average $\langle s_x \rangle$ of a $\Delta p/p_0=10^{-4}$ particle as a function of time, as tuned by the PID controller. It seemingly becomes flat at later times as $k_s$ converges to the optimum value.}
	\label{fig:sx_for_dp_1e_4}
\end{figure}

\begin{figure}
	\centering
	\includegraphics[width=1\linewidth]{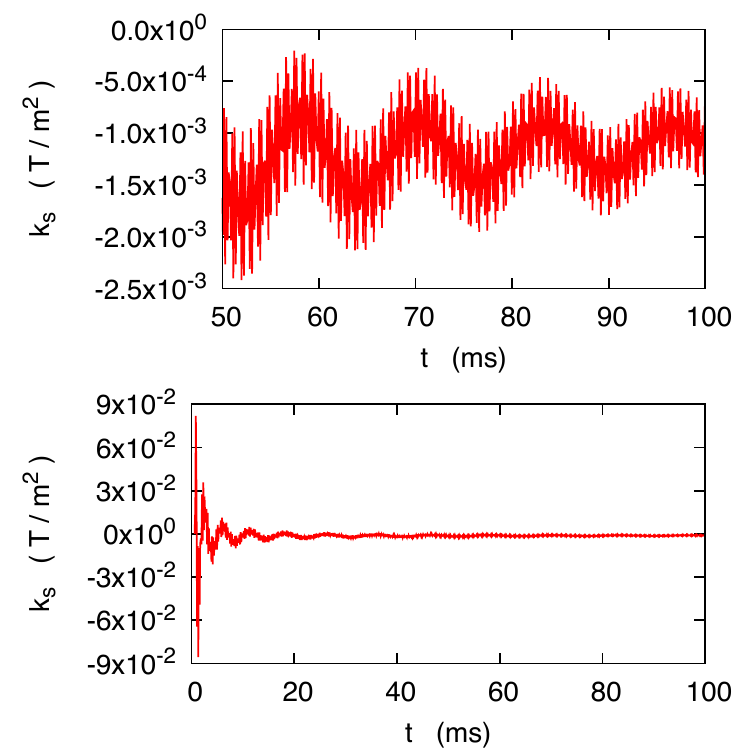}
	\caption{PID tuning of the sextupole strength corresponding to Figure \ref {fig:sx_for_dp_1e_4}. The top plot focuses on the last 50 ms. $k_s$ converged nearby the optimum value that was obtained by simple scanning and fitting algorithm in Section \ref{sect:fixedsext}.}
	\label{fig:ks_for_dp_1e4}
\end{figure}

Figure \ref{fig:sx_for_dp_1e_4} shows $\langle s_x \rangle$, as simulated after the coefficients were adjusted. The larger oscillation at the beginning of the simulation was partly due to the large synchrotron oscillations and mainly because the initial $k_s$ estimations were far away from the optimum value. Figure \ref{fig:ks_for_dp_1e4} shows the corresponding sextupole strength as a function of time. It took only a few iterations for $k_s$ to converge to the optimum value. It is worth mentioning that the same solution could be achieved with small PID coefficients, yet in longer time. In that case, the oscillation amplitudes of $k_s$ and $\langle s_x \rangle$ also become larger. As will be explained in Section \ref{sect:imperfections}, a longer update period is inevitable with noisy $\langle s_x \rangle$ measurements.

\subsection{Simulations with a beam}\label {sect:beamsim}

In this section, the simulations were conducted with a beam that was represented by 74 particles (Figure \ref{fig:distribution}). The particles were simulated simultaneously, and the average in-plane spin of the beam was periodically calculated from Eq. \ref{eq:avg_sx}. At every period, $k_s$ was estimated via Eq. \ref{eq:pid_formula} for updating the sextupoles.

\begin{figure}
	\centering
	\includegraphics[width=\linewidth]{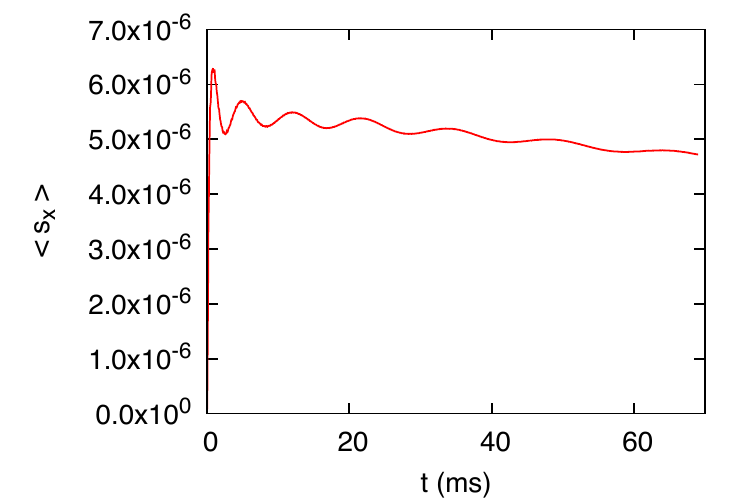}
	\caption{Evolution of $\langle s_x \rangle$ of the beam as a function of time. It drifts by $-1.3 \times 10^{-5}$ rad/s, which is negligible compared to the SCT requirement of the storage ring EDM experiments.}
	\label{fig:sx_after_pid}
\end{figure}

Figure \ref{fig:sx_after_pid} shows the time average of the horizontal spin component of the beam.  The quadrupoles were set to $|k_q|= 0.05$ T/m, which corresponds to $Q_x=1.92$. The drift in the plot could possibly be fixed by a nonzero integral coefficient $K_I$, which was not enabled due to the short simulation time. Nevertheless, $d\langle s_x \rangle/dt=-1.3 \times 10^{-5}$ rad/s is negligible compared to the no-sextupole scenario. 

\begin{figure}
	\centering
	\includegraphics[width=1\linewidth]{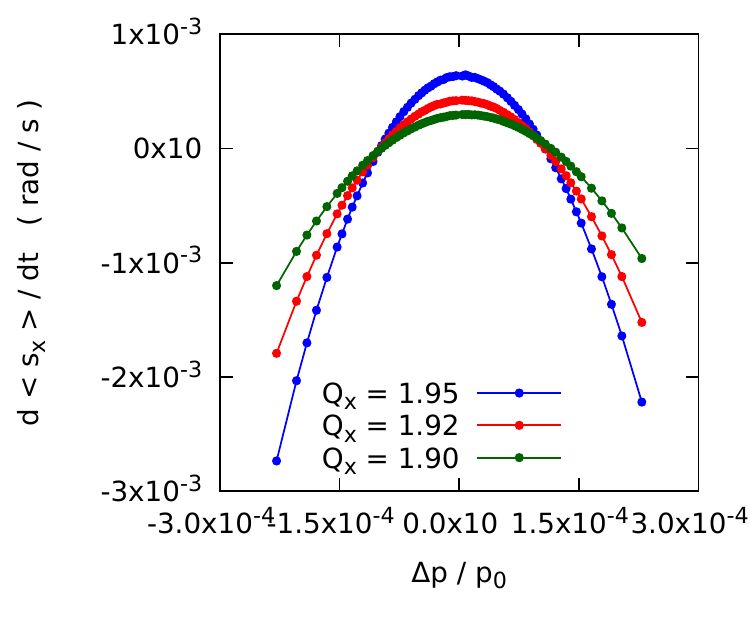}
	\caption{In-plane spin precision rates for different $Q_x$ values as a function of momentum spread. The simulations were made with 74 particles with the distribution shown in Figure \ref{fig:distribution}. The sextupoles were tuned by a PID controller. Different colors correspond to different quadrupole strengths.}
	\label{fig:wa_vs_dp_p_after_pid}
\end{figure}

Figure \ref{fig:wa_vs_dp_p_after_pid} shows the average spin precession rate as a function of the momentum spread for three different quadrupole configurations: $|k_q|=0.04$ T/m, $|k_q|=0.05$ T/m and $|k_q|=0.06$ T/m, with horizontal tunes of 1.90, 1.92 and 1.95, respectively. While the spin precession tends to spread more for the higher $Q_x$ cases, the overall variation in the SCT is acceptable. This is particularly important for the hybrid ring design \cite{ref:hybrid}, where a number of runs with different betatron tunes were proposed. A comparison with fixed sextupole strength case (Figure \ref{fig:shifts_with_fixed_sextupole}) shows that the effect of the betatron tune shift reduces by an order of magnitude with real-time tuning.

All of the above simulations were made with the same set of PID coefficients: $K_I=0;~K_P=300$; $K_D=300$ s. Several simulations were repeated by varying the last two, keeping the $K_D/K_P$ ratio constant. The simulations with smaller coefficients yielded similar $\langle s_x \rangle$ and $k_s$ values at the end with a longer convergence time. On the other hand, too large coefficients caused unstable motion. In conclusion, the PID controller does not require a precise adjustment. It is also worth emphasizing that no initial estimation of sextupole strength was required in the simulations. 

\subsection{Transverse motion}

The RF cavity forces every particle to have a fixed average revolution time around the ring. As mentioned above, setting the revolution frequency to that of the design particle (magic momentum with zero betatron oscillations) helps canceling the in-plane spin precession to first order. However, the betatron oscillations cause a lag at each revolution, which is compensated by the RF cavity by changing the momentum, leading to in-plane spin precession. 

The beam in the previous simulations was injected through the design orbit with no transverse offset. Therefore, the forced betatron oscillations on the horizontal plane were coupled to the momentum spread, and their effect on $\langle s_x \rangle$ was reduced with no extra effort by the sextupoles. 

\begin{figure}
	\centering
	\includegraphics[width=0.9\linewidth]{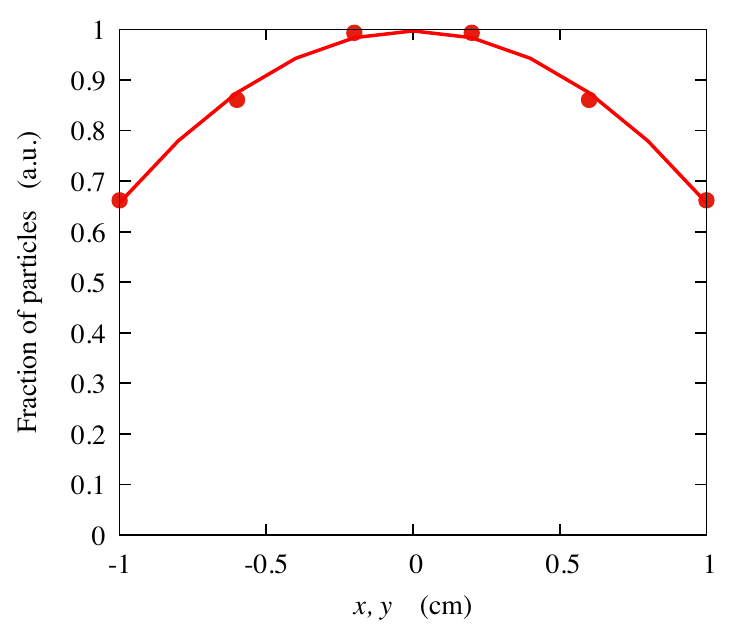}
	\caption{The initial positions of the particles followed this elliptical distribution. While only six $x,~y$ values are depicted in this plot, the simulated particles were distributed at randomly chosen positions according to the probability curve.}
	\label{fig:x_hist}
\end{figure}

\begin{figure*}
	\centering
	\includegraphics[width=1\textwidth]{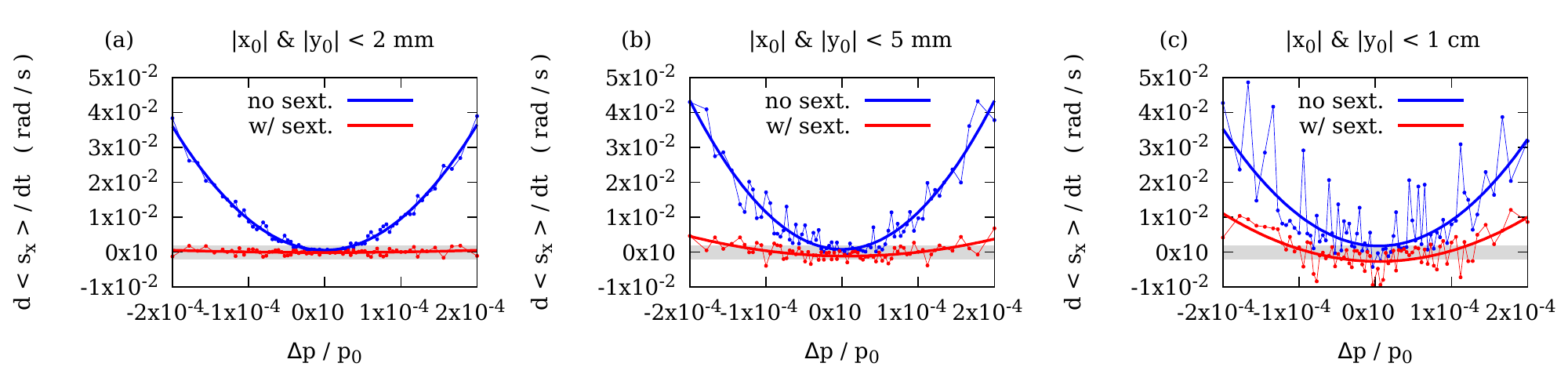}
	\caption{In-plane spin precession rates with respect to the momentum spread for various beam radii of: smaller than (a) 2 mm , (b) smaller than 5 mm, and (c) smaller than 1 cm. The thick curves are second order polynomial fits to the data. The gray area covers $\pm 2$ mrad/s as a reference. As expected, the sextupoles perform better with smaller beams.}
	\label{fig:comparison}
\end{figure*}

With the additional two degrees of freedom (by free horizontal and vertical betatron oscillations), a sextupole strength usually has competing effects on these three terms. In this section, the particles in the beam were given random offsets within an elliptical distribution (Figure \ref{fig:x_hist}). Figure \ref{fig:comparison} shows a comparison of the in-plane spin precession rate of various beam sizes. Spin precession rate is slowed down by one to two orders of magnitude for the smallest beam (a), and up to ten times for the 5 mm radius beam (b). The improvement becomes less significant for the 1 cm radius beam (c). Therefore, a careful lattice design \cite{ref:yuri_sext} and/or a more complex real-time tuning algorithm become essential to achieve a long SCT for large beams. A more complex sextupole system could solve the problem as well, but this is beyond the scope of this work.

\begin{figure}
	\centering
	\includegraphics[width=\linewidth]{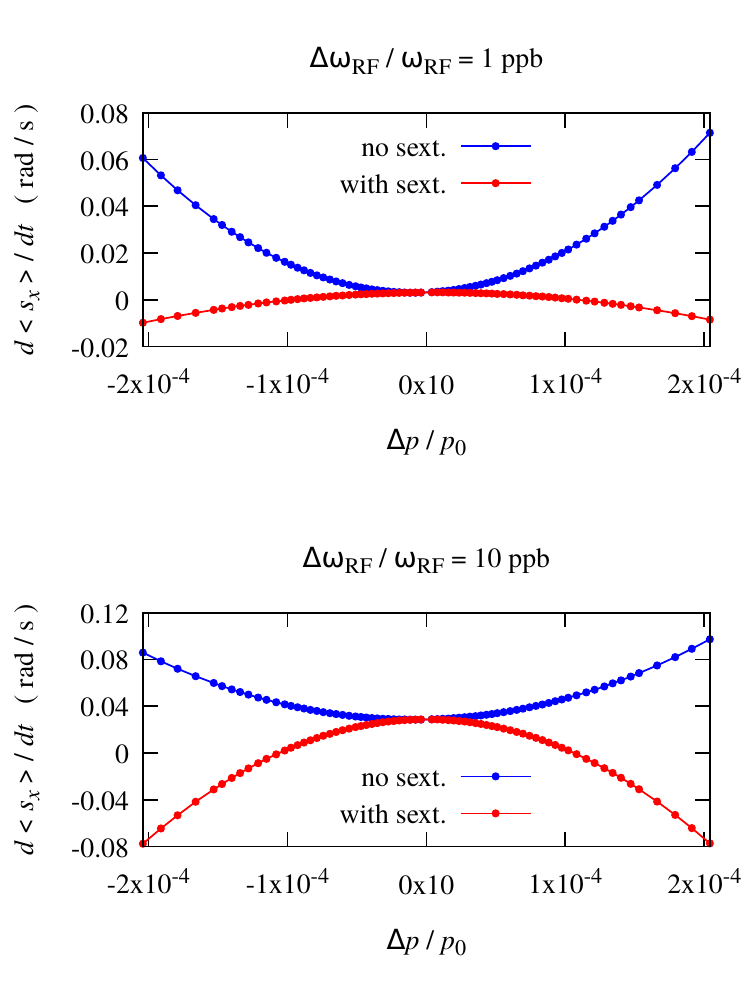}
	\caption{Spin precession rate vs. momentum spread for two RF frequency offsets. As the RF cavity fixes the first order effect ($\Delta p/p_0$), $\Delta \omega_{RF}/\omega_{RF}$ causes a reduction in spin precession rate regardless of the sextupoles. Despite the reduced performance, the real-time tuned sextupoles improve the spin precession rate significantly at 1 ppb offset (see Figure \ref{fig:comparison}(a)). They turn out to be comparable to no-sextupole case at 10 ppb offset.}
	\label{fig:spt_vs_dp_p_vs_RF_freq}
\end{figure}

\section{Effect of imperfections}\label {sect:imperfections}
The performance of this method is limited by several imperfections. Among the field perturbations, vertical magnetic field is the most limiting factor. It can both degrade the sextupole performance and cause an ambiguity regarding the average spin value. External electric fields at practical levels do not have a significant effect on $s_x$, as long as the RF cavity fixes the energy drifts properly. Measurement and feedback errors can degrade the controller performance too. This section is devoted to the most dominant of these imperfections. The simulations were made with the same momentum spread as in the previous sections, with the particles injected at $x_0=y_0=0$ unless stated otherwise.  

\subsection{Shift in the RF frequency}
The parenthesis in Eq. \ref{eq:dsx_dt} can be canceled to first order by using the frozen spin method and beam bunching, even if the beam is not injected at the design orbit. However, a frequency offset $\Delta \omega_{RF}/\omega_{RF}$ in the RF cavity shifts the energy and the radial position of the beam. A set of simulations were made with and without the sextupoles, using an off-frequency RF cavity. As shown in Figure \ref{fig:spt_vs_dp_p_vs_RF_freq}, insufficient cancellation of the first order effect ($\Delta p/p_0$) results in increased spin precession rate with the RF frequency offset. The real-time tuned sextupoles improve it by another order of magnitude for 1 ppb offset, but they do not help much when the RF shift is increased to 10 ppb. 1 ppb offset was studied with a 5 mm radius beam as well, and the result was comparable to the tiny beam case.

\subsection{External magnetic field}
A vertical magnetic field can affect the in-plane spin precession in two ways. As seen in the first term of Eq. \ref{eq:bmt}, for a longitudinally polarized beam, it can couple with $G$, and change the in-plane spin precession rate. Secondly, it shifts the beam on the horizontal plane, resulting in an energy shift in the deflectors. This also changes the in-plane spin precession rate according to Eq. \ref{eq:dsx_dt}. Eventually, the sextupoles under-perform by these two effects.

\begin{figure}
\centering
\includegraphics[width=\linewidth]{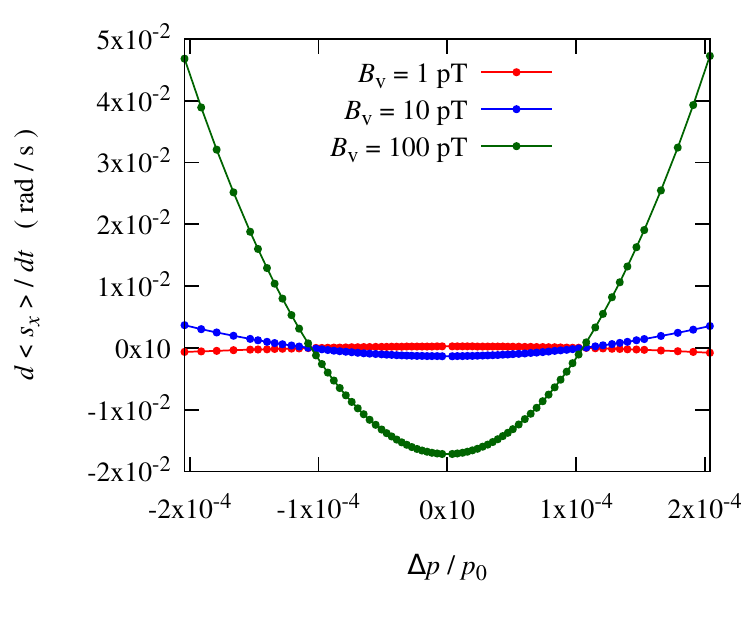}
\caption{Spin precession rate as a function of momentum spread for various uniform vertical magnetic field cases. Comparison with Figure \ref{fig:comparison}(a) shows that $B_V=1$ pT causes a slight degradation at small, and no significant change at large $\Delta p/p_0$ values. The PID still performs well at $B_V=10$ pT, while it is not beneficial any more at $B_V=100$ pT level field.}
\label{fig:spt_vs_dp_p_with_Bver}
\end{figure}

Figure \ref{fig:spt_vs_dp_p_with_Bver} shows simulation results for several uniform vertical magnetic field cases. As compared with Figure \ref{fig:comparison}(a), 1 pT level magnetic field had a negligible effect on the spin precession rates. While 10 pT field had a visible effect, the overall spin precession rate was still an order of magnitude slower than the no-sextupole case. However, the PID-controlled sextupoles were not beneficial any more with 100 pT level average vertical magnetic field. Varying magnetic field was also simulated by changing it randomly between 0 and 20 pT at every $k_s$ update period. As a result, the spin precession was almost identical to the fixed 10 pT field case. Additional simulations with a  5 mm radius beam showed that 10 pT vertical magnetic field did not have a significant effect (as compared with Figure \ref{fig:comparison}(b)).

\subsection{Nonlinearities, hysteresis, and ripples}
\begin{figure}
	\centering
	\includegraphics[width=\linewidth]{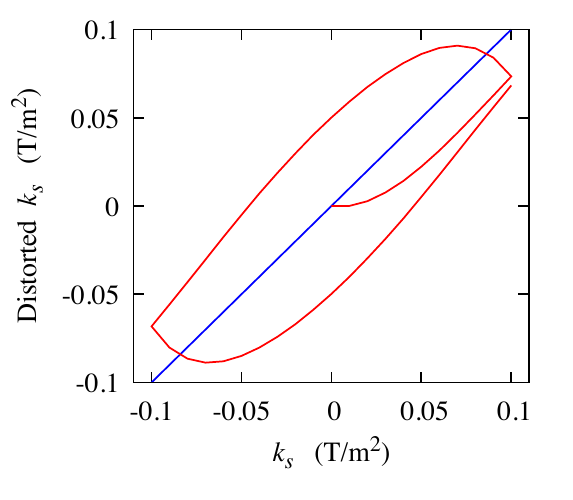}
	\caption{Distorted $k_s$ vs. the set value. The blue line represents the no-hysteresis case, where the generated field corresponds to the set value. The red curve represents the hysteresis of the magnetic field, and determines the envelope of the distorted $k_s$.}
	\label{fig:hysteresis}
\end{figure}

As mentioned above, $k_s$ in the previous simulations was estimated by Eq. \ref{eq:pid_formula}. However,  the sextupole magnetic field may have a hysteresis in presence of a ferromagnetic core. A set of simulations were conducted by introducing this effect in the $k_s$ estimations. The envelope of the hysteresis-like curve is drawn in Figure \ref{fig:hysteresis}. The envelope was chosen to have a large nonlinearity at the extreme $k_s$ values. The simulations revealed no significant deviation from the no-distortion cases. This originates from the fact that after a number of cycles, the distorted $k_s$ averages to the ideal value. 

Power supply ripples have a similar effect with the magnetic hysteresis. Figure \ref{fig:ks_for_dp_1e4} shows that the amplitude of $k_s$ ranges between 0.002 T/m$^2$ and 0.1 T/m$^2$. Based on these values, ripples were introduced by modulating $k_s$ by $5\times 10^{-4}~\text{T/m}^2$ amplitude and 5 kHz frequency. Simulation results yielded no significant difference from the no-ripple case.

\subsection{Polarimeter noise}
As explained in Section \ref{sect:varsext}, real-time sextupole tuning relies on periodic measurement of the time average $\langle s_x \rangle$ and updating $k_s$ accordingly. Because of the time averaging, white noise contributes less at later times to the first and second terms of Eq. \ref{eq:pid_formula}. On the other hand, regardless of the averaging time, the noise at the last measurement is always reflected directly in $\langle s_x \rangle ^\text{diff}$, introducing a strong sensitivity to the measurement noise. Therefore, $K_D$ should be zero for noisy measurements.

In a set of simulations, a random noise was added to the $s_x(t)$ estimations at every PID update period $\tau$. For 100 $\mu$rad noise, no stable result was achieved unless $K_P$ was reduced to 5 (as opposed to 300). Overall, the simulations showed that the effect of the statistical noise can be eliminated by setting $K_D=0$, and decreasing $K_P$. The update period of $k_s$ can also be increased for a better statistics. Then, because of the long response time, $\langle s_x \rangle$ makes larger oscillations. As shown in Figure \ref{fig:sx_for_dp_1e_4}, there is a room for several orders of magnitude larger and slower oscillations. 

Obviously, this solution is useful as long as the storage time permits.  For the storage ring EDM experiments, a $10^{-29}~e \cdot$cm sensitivity corresponds to a spin precession rate of approximately $1-10$ nrad/s in $10^7$ seconds. Assuming $10^4$ injections, the sensitivity of the polarimeter is expected to be $0.1-1~\mu$rad per injection, which translates to a sub-milliradian level statistical sensitivity per second. Then, using a sufficiently small $K_P$, and updating the PID controller at every second, a stable solution can be reached within several hundred seconds.
 
\section{Summary and Conclusions}
Sextupoles can be utilized in storage ring spin physics experiments to correct the second order chromaticity-related effects and prolong the SCT to the required level. However, its optimization is a time consuming process and slight changes in the ring parameters affect the SCT significantly. Therefore, a real-time sextupole tuning may be inevitable.

The storage ring in this work was perfectly symmetric with one family of quadrupoles and sextupoles, which had alternating polarities at every straight section. A more complex storage ring could provide a better SCT, while requiring a more complex PID controller (with features like adaptivity, limiting the maximum sextupole strength, etc.) for each sextupole family. 

For the best performance, the average vertical magnetic field should be kept at 1 pT level and the RF frequency should be accurate within 1 ppb or better. Hysteresis, nonlinearities and ripples that modify the sextupole fields do not have a significant effect on the PID performance.

The simulation time in this study was kept at milliseconds scale because of limited CPU speed, as well as the limitations of the simulation tool. Therefore, the PID controller was updated at every 0.1 ms. On the other hand, the simulations showed that the same stable solutions could be achieved by slower controller updates and smaller feedback coefficients. This is especially necessary to handle large polarimeter noise. Extrapolating the simulation results, it was concluded that a milliradian level statistical sensitivity from the polarimeters will be harmless to the real-time tuning operation.

Besides real-time tuning, a feedback-based optimization can be considered as a tool to optimize the sextupole strength as well. The sextupoles can be tuned by a PID controller in a test run after the experimental conditions change. Then, the following storages can be performed with fixed sextupoles using the optimized strength.

\section*{Acknowledgments}
This work was supported by IBS-R017-D1 of the Republic of Korea. I would like to thank Yannis K. Semertzidis and SungWoo Youn for helpful discussions.

\end{document}